\documentclass[preprint,tightenlines,showpacs,amsmath,amssymb]{revtex4}

\usepackage{epsfig,epsf,graphics,psfrag}
\usepackage{times}
\usepackage{float}
\usepackage{color}
\usepackage{amsfonts,amssymb,stmaryrd,latexsym,amsmath}
\usepackage{multirow}
\usepackage{array} 
\usepackage{hhline}
\usepackage{booktabs}
\usepackage{enumerate}
\usepackage{slashed}

\begin{document}

\bibliographystyle{unsrt}

\title{Could the observation of $X(5568)$ be resulted by the near threshold rescattering effects?}

\author{Xiao-Hai Liu$^1$\footnote{liuxh@th.phys.titech.ac.jp} and Gang Li$^{2}$\footnote{gli@mail.qfnu.edu.cn}}

\affiliation{ $^1$Department of Physics, H-27, Tokyo Institute of Technology, Meguro, Tokyo 152-8551, Japan}

\affiliation{$^2$ College of Physics and Engineering, Qufu Normal University, Qufu 273165, China}

\date{\today}

\begin{abstract}

We investigate the invariant mass distributions of $B_s\pi$ via different rescattering processes. Because the triangle singularity (TS) could be present for a very broad incident energy region, it can be expected that the TS peaks may simulate the resonance-like bump $X(5568)$ observed by the D0 collaboration. The highly process-dependent characteristic of TS mechanism offers a criterion to distinguish it from other dynamic mechanisms.


\pacs{~14.40.Rt,~12.39.Mk,~14.40.Nd}

\end{abstract}

\maketitle

\section{Introduction}
 The study on exotic hadron spectroscopy is experiencing a renaissance in the last decade. More and more charmonium-like and bottomonium-like states (called $XYZ$) have been announced by experiments in various processes~(see Refs.~\cite{Brambilla:2010cs,Agashe:2014kda,Esposito:2014rxa,Chen:2016qju} for a review). Several charged structures with a hidden $\bar bb$ or $\bar cc$, such as the$Z_c^\pm(4430)$~\cite{Choi:2007wga,Aaij:2014jqa},
$Z_b^\pm(10610,10650)$~\cite{Belle:2011aa},
$Z_c^\pm(3900)$~\cite{Ablikim:2013mio,Liu:2013dau}, and
$Z_c^\pm(4020)$~\cite{Ablikim:2013emm} were observed by experiments, which would be exotic state candidates. Very recently, the D0 collaboration observed a narrow structure $X(5568)$ in the $B_s^0\pi^\pm$ invariant mass spectrum with $5.1\sigma$ significance~\cite{D0:2016mwd}. The mass and width are measured to be $M_X=5567.8\pm 2.9^{+2.9}_{-1.9}$ MeV and $\Gamma_X=21.9\pm 6.4^{+5.0}_{-2.5}$ MeV, respectively. The quark components of the decaying final state $B_s^0 \pi^\pm$ are $su\bar b \bar d$ (or $sd\bar b \bar u$), which requires $X(5568)$ should be a structure with four different valence quarks.

After the discovery of $X(5568)$, several theoretical investigations are carried out in order to understand its underlying structure. In Ref.~\cite{Agaev:2016mjb}, the mass and decay constant of $X(5568)$ were computed within the two-point sum rule method using the diquark-antidiquark interpolating current. The mass and pole residue are studied with the QCD sum rules in~\cite{Wang:2016mee}. Its mass spectrum was also calculated in Ref.~\cite{Wang:2016tsi,Chen:2016mqt} using diquark-antidiquark type interpolating current. In Ref.~\cite{Xiao:2016mho}, the authors estimated the partial decay width $X(5568) \to B_s^0 \pi^+$ with $X(5568)$ being an $S$-wave $B\bar K$ molecular state.

It is necessary to study some other possibilities, before we claim that $X(5568)$ is a genuine particle, such as tetraquark or molecular state. Some non-resonance explanations have ever been proposed to connect resonance-like peaks with kinematic singularities induced by the rescattering effects ~\cite{Chen:2011pv,Chen:2011zv,Bugg:2011jr,Chen:2011xk,Wang:2011yh,Wu:2011yx,Wang:2013cya,Liu:2013vfa,Liu:2014spa,Szczepaniak:2015eza,Liu:2015fea,Liu:2015cah,Liu:2016dli}. It is shown that sometimes it is not necessary to introduce a genuine resonance to describe a resonance-like structure, because some kinematic singularities of the rescattering amplitudes will behave themselves as bumps in the invariant mass distributions. In this work, we are trying to use the so-called triangle singularity (TS) mechanism to describe the observation of $X(5568)$.

The work is organized as follows: In Section II, the TS mechanism is briefly introduced; In Section III, we discuss several rescattering processes where the TS can be present and simulate the resonance. A brief summary is given in Section IV.

\section{TS mechanism}

\begin{figure}[b]
	\centering
	\includegraphics[width=0.3\hsize]{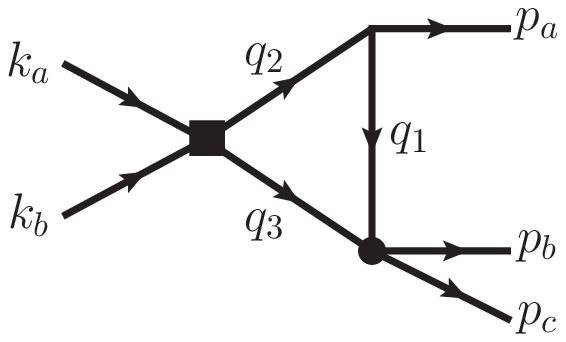}
	\caption{Triangle rescattering diagram under discussion. The internal mass which corresponds to the internal momentum $q_i$ is $m_i$ ($i$$=$1, 2, 3). The momentum symbols also represent the corresponding particles.}\label{TSmechanism}
\end{figure}

 The possible manifestation of the TS was first noticed in 1960s. It is found that the TS's of rescattering amplitudes can mimic resonance structures in the corresponding invariant mass distributions~\cite{Peierls:1961zz,Goebel:1964zz,hwa:1963aa,landshoff:1962aa,Aitchison:1969tq,Aitchison:1964ak,coleman:1965aa,bronzan:1964aa,fronsdal:1964aa,norton:1964aa,schmid:1967aa}. This offers a non-resonance explanation about the resonance-like peaks observed in experiments. Unfortunately, most of the proposed cases in 1960s were lack of experimental support. The TS mechanism was rediscovered in recent years and used to interpret some exotic phenomena, such as the largely isospin-violation processes, the production of some exotic states and so on~\cite{Wu:2011yx,Ketzer:2015tqa,Wang:2013cya,Liu:2013vfa,Liu:2014spa,Szczepaniak:2015eza,Guo:2014iya,Liu:2015taa,Liu:2015cah,Liu:2016dli,Liu:2015fea,Guo:2015umn}.
 
 For the triangle diagram in Fig.~\ref{TSmechanism}, all of the three internal lines can be on-shell simultaneously under some special kinematic configurations. This case corresponds to the leading Landau singularity of the triangle diagram, and this leading Landau singularity is usually called the TS. The physical picture concerning the TS mechanism can be understood like this: The initial particles $k_a$ and $k_b$ firstly scatter into particles $q_2$ and $q_3$, then the particle $q_1$ emitted from $q_2$ catches up with $q_3$, and finally $q_2$ and $q_3$ will rescatter into particles $p_b$ and $p_c$. This implies that the rescattering diagram can be interpreted as a classical process in space-time with the presence of TS, and the TS will be located on the physical boundary of the rescattering amplitude~\cite{coleman:1965aa}. 
 
 The TS mechanism is very sensitive to the kinematic configurations of rescattering diagrams. It is therefore necessary to determine in which kinematic region the TS can be present.
 In Fig.~\ref{TSmechanism}, we define the invariants $s_1=(k_a+k_b)^2$, $s_2=(p_b+p_c)^2$ and $s_3=p_a^2$. The locations of TS can be determined by solving the so-called Landau equations \cite{Landau:1959fi,Eden:1966,bonnevay:1961aa}. For the diagram in Fig.~\ref{TSmechanism},
if we fix the internal masses $m_i$, the external invariants $s_2$ and $s_3$, we can obtain the solutions of TS in $s_1$, i.e.,
\begin{eqnarray}
s_1^{\pm}&=&(m_2+m_3)^2+\frac{1}{2m_1^2} {\LARGE[}(m_1^2+m_2^2-s_3)(s_2-m_1^2-m_3^2)-4m_1^2 m_2 m_3 \nonumber \\  
&\pm& \lambda^{1/2}(s_2, m_1^2, m_3^2)\lambda^{1/2}(s_3,m_1^2,m_2^2){\LARGE ]},
\end{eqnarray}
with $\lambda(x,y,z)= (x-y-z)^2- 4yz$.
Likewise, by fixing $m_i$, $s_1$ and $s_3$ we can obtain the similar solutions of TS in $s_2$, i.e.,
\begin{eqnarray}
s_2^{\pm}&=&(m_1+m_3)^2+\frac{1}{2m_2^2} {\LARGE[}(m_1^2+m_2^2-s_3)(s_1-m_2^2-m_3^2)-4m_2^2 m_1 m_3 \nonumber \\  &\pm& \lambda^{1/2}(s_1,  m_2^2,  m_3^2)\lambda^{1/2}(s_3,m_1^2,m_2^2){\LARGE ]}.
\end{eqnarray}
By means of the single dispersion representation of the 3-point function, we learn that within the physical boundary only the solution $s_1^-$ or $s_2^-$ will correspond to the TS of rescattering amplitude, and $\sqrt{s_1^-}$ and $\sqrt{s_2^-}$ are usually defined as the anomalous thresholds \cite{Liu:2015taa,Eden:1966,bonnevay:1961aa}. For convenience, we further define the normal threshold $\sqrt{s_{1N}}$ ($\sqrt{s_{2N}}$) and the critical value $\sqrt{s_{1C}}$ ($\sqrt{s_{2C}}$) for $s_1$ ($s_2$) as follows \cite{Liu:2015taa}, 
\begin{eqnarray}\label{s1Ns1C}
&& s_{1N}=(m_2+m_3)^2,\ s_{1C}=(m_2+m_3)^2 +\frac{m_3}{m_1}[(m_2-m_1)^2-s_3], \nonumber \\
&& s_{2N}=(m_1+m_3)^2,\ s_{2C}=(m_1+m_3)^2 +\frac{m_3}{m_2}[(m_2-m_1)^2-s_3].
\end{eqnarray}
If we fix $s_3$ and the internal masses $m_{1,2,3}$, when $s_1$ increases from $s_{1N}$ to $s_{1C}$, $s_2^-$ will move from $s_{2C}$ to $s_{2N}$. Likewise, when $s_2$ increases from $s_{2N}$ to $s_{2C}$, $s_1^-$ will move from $s_{1C}$ to $s_{1N}$. This is the kinematic region where the TS can be present.
The discrepancies between normal and anomalous thresholds can also be used to represent the TS kinematic region. The maximum values of these discrepancies take the form
\begin{eqnarray} \label{deltas1s2}
\Delta_{s_1}^{\max}&=&\sqrt{s_{1C}} - \sqrt{s_{1N}}\approx
\frac{m_3}{2m_1(m_2+m_3)}[(m_2-m_1)^2-s_3], \nonumber \\
\Delta_{s_2}^{\max}&=&\sqrt{s_{2C}} - \sqrt{s_{2N}}\approx
\frac{m_3}{2m_2(m_1+m_3)}[(m_2-m_1)^2-s_3].
\end{eqnarray} 

In Ref.~\cite{schmid:1967aa}, it was argued that for the single channel rescattering process, when the corresponding resonance-production tree diagram is added coherently to the triangle rescattering diagram, the effect of the triangle diagram is nothing more than a multiplication of the singularity from the tree diagram by a phase factor. Therefore the singularities of triangle diagram cannot produce obvious peaks in the Dalitz plot projections. This is the so-called Schmid theorem. But for the coupled-channel cases, the situation will be quite different from the single channel case discussed in Ref.~\cite{schmid:1967aa}. For the rescattering diagrams which will be studied in this paper, the intermediate and final states are different, therefore the singularities induced by the rescattering processes are still expected to be visible in the Dalitz plot projections. The reader is referred to Refs.~\cite{Aitchison:1969tq,Goebel:1982yb} for some comments about the Schmid theorem, and Refs.~\cite{Anisovich:1995ab,Szczepaniak:2015hya} for some discussions about the coupled-channel case. We will focus on the coupled-channel cases in this work.

\section{ Production of $B_s\pi$ via rescattering processes}



\subsection{Triangle diagram}
\begin{figure}[htb]
	\centering
	\includegraphics[width=0.53\hsize]{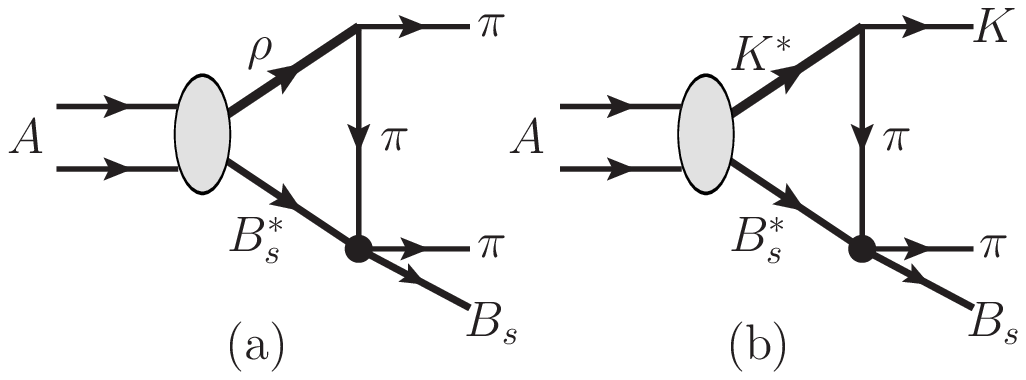}
	\caption{Production of $B_s\pi$ via the triangle rescattering diagrams (a) $[B_s^*\rho\pi]$-loop and (b) $[B_s^* K^*\pi]$-loop. $A$ represents the incident state.}\label{feymanTri}
\end{figure}

The mass of $X(5568)$ is very close to the $B_s^*\pi^\pm$-threshold, which is about 5555 MeV~\cite{Agashe:2014kda}. One may wonder whether there are some connections between $X(5568)$ and the coupled-channel scattering $B_s^*\pi\to B_s\pi$ near threshold. In the high energy collisions, the production of $B_s\pi$ may receive contributions from rescattering diagrams illustrated in Fig.~\ref{feymanTri}. The intriguing characteristic of this kind of diagrams is TS's are expected to be present in the rescattering amplitudes, which may result in some resonance-like bumps in $B_s\pi$ distributions around the $B_s^*\pi$-threshold accordingly. 

\begin{table}\caption{TS kinematic region corresponding to the rescattering diagrams in Fig.~\ref{feymanTri}, in unit of GeV.} 
	\begin{center}
		\begin{tabular}{c|c|c|c|c|c|c}
			\hline
			Diagram & $ \sqrt{s_{1N}} $ & $ \sqrt{s_{1C}} $ & $ \Delta _{s_1}^{\max } $ & $ \sqrt{s_{2N}} $ & $ \sqrt{s_{2C}} $ & $ \Delta _{s_2}^{\max } $ \\
			\hline
			(a)&6.191 & 7.297 & 1.106 & 5.555 & 5.792 & 0.237 \\
			\hline
			(b)&6.311 & 7.250 & 0.939 & 5.555 & 5.731 & 0.176 \\
			\hline
		\end{tabular}
	\end{center}\label{KMregion}
\end{table} 
The momentum and invariants conventions of Fig.~\ref{feymanTri} are the same with those of Fig.~\ref{TSmechanism}.
According to Eqs.~(\ref{s1Ns1C}) and (\ref{deltas1s2}), the kinematic region where the TS can be present is displayed in Table~\ref{KMregion}. It can be seen that the kinematic region of TS in $s_1$ is very large for both of the diagrams in Fig.~\ref{feymanTri}. $\Delta_{s_1}^{\max}$ is nearly 1 GeV for each of the diagrams. Firstly, this is because the quantity $[(m_2-m_1)^2-s_3]$ in Eq.~(\ref{deltas1s2}) is large. Physically, this quantity corresponds to the phase-space factor for $\rho\to\pi\pi$ ($K^*\to K\pi$), which is sizable. Secondly, the ratio $m_3/m_1$ is equal to $M_{B_s^*}/M_\pi$, which is also quite large. 
This means that the kinematic conditions of the presence of TS can be fulfilled in a very broad energy region of incident states.
The kinematic requirement on the incident state would be largely relaxed, which is an advantage to observe the effects resulted by the TS mechanism. On the other hand, the kinematic region of TS in $s_2$ is relatively smaller. $\Delta_{s_2}^{\max}$ is about 0.2 GeV for each of the diagrams, which implies that the TS peaks in $B_s\pi$ distributions may not stay far away from the $B_s^*\pi$-threshold (normal threshold $\sqrt{s_{2N}}$).

We will naively construct some effective Lagrangians to estimate the behaviour of the rescattering amplitudes. Taking into account the conservation of angular momentum and parity, the quantum numbers of the incident state $A$ are set to be $J^P$=$1^+$. Some of the Lagrangians read
\begin{eqnarray}
\mathcal{L}_{AB_s^* V}&=& g_{A} \epsilon^{\mu\nu\alpha\beta} \partial_\mu A_\nu V_\alpha {B^*_s}_\beta , \\
\mathcal{L}_{VPP}&=& i g_{VPP} V^\mu [P \partial_\mu P^\prime, \partial_\mu P P^\prime],
\end{eqnarray}
where $V$ and $P^{(\prime)}$ represent the light vector and pesudoscalar mesons respectively. 
The process $B_s^*\pi\to B_s\pi$ can be a $P$-wave scattering, and the corresponding Lagrangian takes the form
\begin{eqnarray}\label{contact}
\mathcal{L}_{B_s^*B_s\pi\pi}&=& g_{CT} \epsilon^{\mu\nu\alpha\beta} {B^*_s}_\mu \partial_\nu B_s \partial_\alpha \pi \partial_\beta \pi. 
\end{eqnarray}
The $P$-wave scattering implies that the quantum numbers of $B_s^*\pi$ and $B_s\pi$ systems would be $J^P$=$1^-$.  
It should be mentioned that the processes $A\to B_s^* V$ in Fig.~\ref{feymanTri} can also happen through weak interactions in the high energy collisions, therefore the parity does not have to be conserved for this vertex.

\begin{figure}[htb]
	\centering
	\includegraphics[width=0.38\hsize]{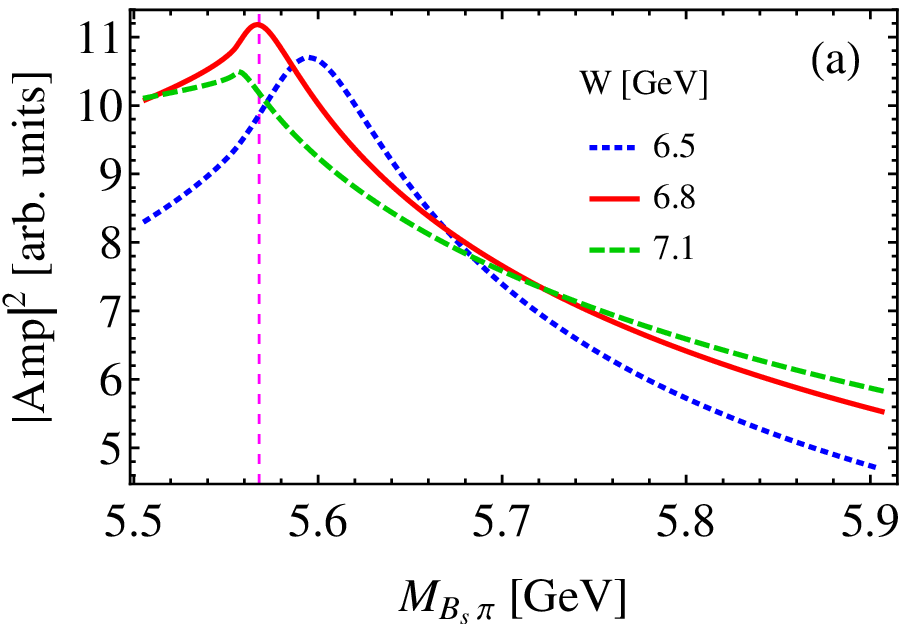}
	\includegraphics[width=0.38\hsize]{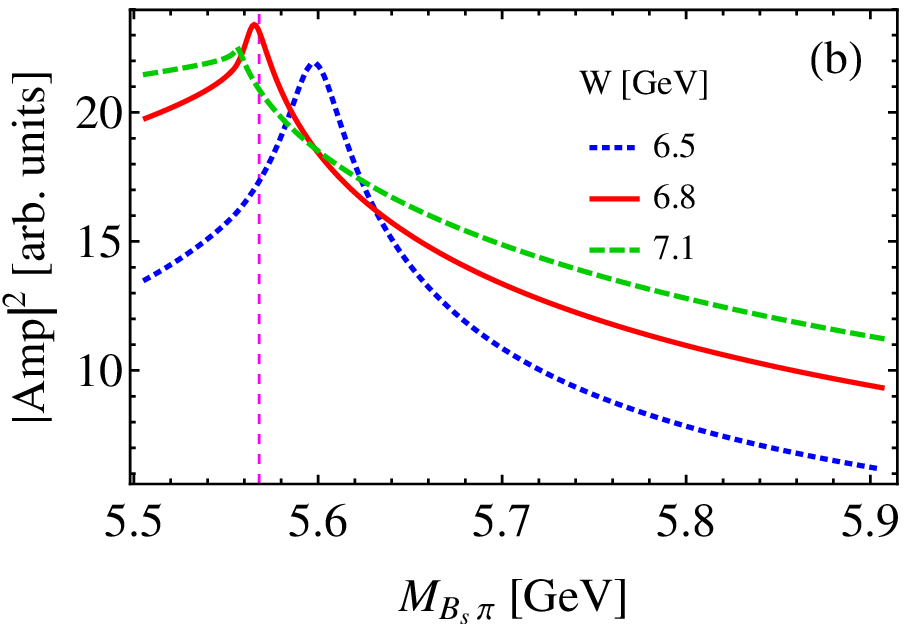}
	\caption{$M_{B_s\pi}$-dependence of the rescattering amplitude squared $|\mbox{Amp}|^2$. Plots (a) and (b) correspond to Figs.~\ref{feymanTri}(a) and (b), respectively. $W$ is the invariant mass $\sqrt{s_1}$ of incident state $A$. The vertical dashed line indicates the position of $X(5568)$.}\label{distributionTri}
\end{figure}

 When the kinematic conditions of TS are fulfilled, it will imply that the particle $q_2$ in Fig.~\ref{TSmechanism} will be unstable. We then introduce a Breit-Wigner type propagator $[q_2^2-m_2^2+i m_2 \Gamma_2]^{-1}$ to account for the width effect when calculating the triangle loop integrals. The complex mass of the intermediate state will remove the TS from physical boundary by a distance~\cite{Aitchison:1964ak}. If the width $\Gamma_2$ is not very large, the TS will lie close to the physical boundary, and the scattering amplitude can still feel the influence of the singularity.
 
 The numerical results for $B_s\pi$ invariant mass distributions corresponding to the rescattering processes in Fig.~\ref{feymanTri} are displayed in Fig.~\ref{distributionTri}. We ignore the explicit couplings but just focus on the line-shape behavior here. The distributions are calculated at several incident energy points. From Fig.~\ref{distributionTri}, it can be seen that some bumps arise around the position of $X(5568)$. Since the bumps around $X(5568)$ can be present for a very broad incident energy region, it is possible that the observation of $X(5568)$ is due to some kind of accumulative effects of the rescattering amplitudes at different incident energies. 
 
 The bumps in Fig.~\ref{distributionTri}(a) are broader compared with those in Fig.~\ref{distributionTri}(b). This is because the decay width of $\rho$-meson (149 MeV) is larger than that of $K^*$-meson (50 MeV) \cite{Agashe:2014kda}. The larger decay width will remove the corresponding TS in the complex-plane further away from the physical boundary. The $P$-wave scattering $B_s^*\pi\to B_s\pi$ will also smooth the TS peaks to some extent.

\subsection{Long range interaction and box diagram}
\begin{figure}[htb]
	\centering
	\includegraphics[width=0.45\hsize]{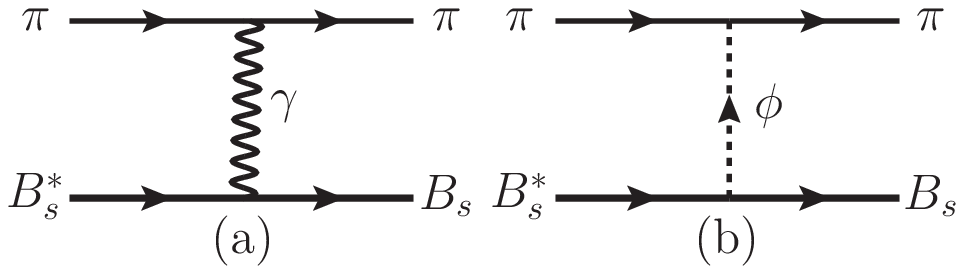}
	\caption{$t$-channel contributions for $B_s^*\pi\to B_s\pi$. (a): photon-exchange; (b): $\phi$-exchange.}\label{EMphi}
\end{figure}

\begin{figure}[htb]
	\centering
	\includegraphics[width=0.38\hsize]{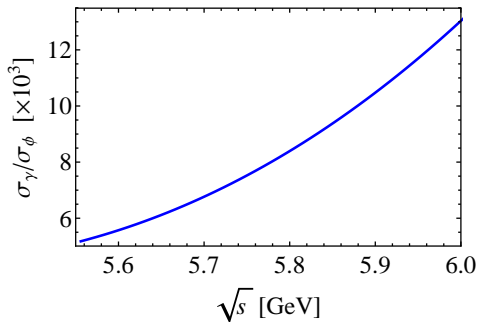}
	\caption{Total cross section ratio $\frac{\sigma_\gamma}{\sigma_\phi}$ corresponding to the $t$-channel scatterings in Fig.~\ref{EMphi}. $\sqrt{s}$ represents the scattering energy.}\label{emratio}
\end{figure}

The scattering process $B_s^*\pi\to B_s\pi$ would be OZI suppressed. In Eq.~(\ref{contact}), we assume a contact interaction, which may account for the short range part of this interaction. If we take into account the $t$-channel contributions, because the momentum transfer in this process will be very small, some long range interactions, such as the electromagnetic (EM) interaction, may become important.
To judge whether the EM interaction may play a role in $B_s^*\pi\to B_s\pi$, we will compare the contributions of $t$-channel processes illustrated in Fig.~\ref{EMphi}. 

In Figs.~\ref{EMphi}(a) and (b), we use the photon- and $\phi$-exchange diagrams to partly account for the EM and strong interactions, respectively. Some effective Lagrangians are constructed as follows
\begin{eqnarray}\label{eminteraction}
\mathcal{L}_{B_s^*B_s\gamma}&=& i g_{B_s^* B_s\gamma} \epsilon^{\mu\nu\alpha\beta} \partial_\mu{B^*_s}_\nu \partial_\alpha \mathcal{A}_\beta B_s , \\
\mathcal{L}_{B_s^*B_s\phi}&=& i g_{B_s^* B_s\phi} \epsilon^{\mu\nu\alpha\beta} \partial_\mu{B^*_s}_\nu \partial_\alpha \phi_\beta B_s , \\
\mathcal{L}_{\gamma\pi\pi}&=& -i e \mathcal{A}^\mu (\partial_\mu \pi^+ \pi^- - \pi^+ \partial_\mu  \pi^-).
\end{eqnarray}
We firstly compare the coupling constants of these two diagrams.
 If adopting the vector meson dominance model \cite{Bauer:1975bw,Casalbuoni:1996pg,Li:2008xm,Zhang:2008ab} , the ratio $R_{\gamma/\phi}\equiv e g_{B_s^* B_s\gamma}/ g_{\phi\pi\pi} g_{B_s^* B_s\phi}$ would be equal to $\sqrt{4\pi \alpha_e}g_{\gamma\phi} / g_{\phi\pi\pi}$, where the couplings $g_{\gamma\phi}$ and $g_{\phi\pi\pi}$ are estimated to be $ 0.0226 $ and $ 0.0072 $ according to the decay widths of $\phi\to e^+e^-$ and $\phi\to \pi\pi$, respectively. The ratio $R_{\gamma/\phi}$ is then obtained to be about $ 0.9 $. 
According to this naive estimation, we can see that the EM couplings may not be very smaller compared with the OZI suppressed strong couplings. Without introducing any form factors to account for the off-shell effects, we further integrate over the moment transfer $t$ and obtain the cross section ratios $\sigma_{\gamma}/\sigma_{\phi}$ for different scattering energies, which is displayed in Fig.~\ref{emratio}. We can see that the cross section corresponding to Fig.~\ref{EMphi}(a) is larger than that corresponding to  Fig.~\ref{EMphi}(b) by about three orders of magnitude. This is mainly because the quantity $1/t$ is much larger than $1/(t-m_\phi^2)$. 
We can then make a quantitative judgment that the contribution of EM interaction may be comparable with that of strong interaction. 

\begin{figure}[htb]
	\centering
	\includegraphics[width=0.6\hsize]{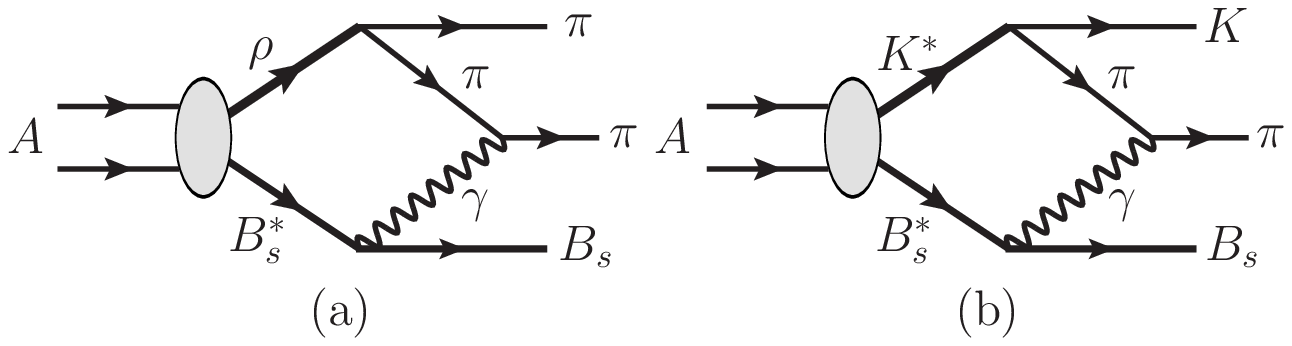}
	\caption{Production of $B_s\pi$ via the box rescattering diagrams (a) $[B_s^*\rho\pi\gamma]$-loop and (b) $[B_s^* K^*\pi\gamma]$-loop.}\label{feynmanBox}
\end{figure}

\begin{figure}[htb]
	\centering
	\includegraphics[width=0.38\hsize]{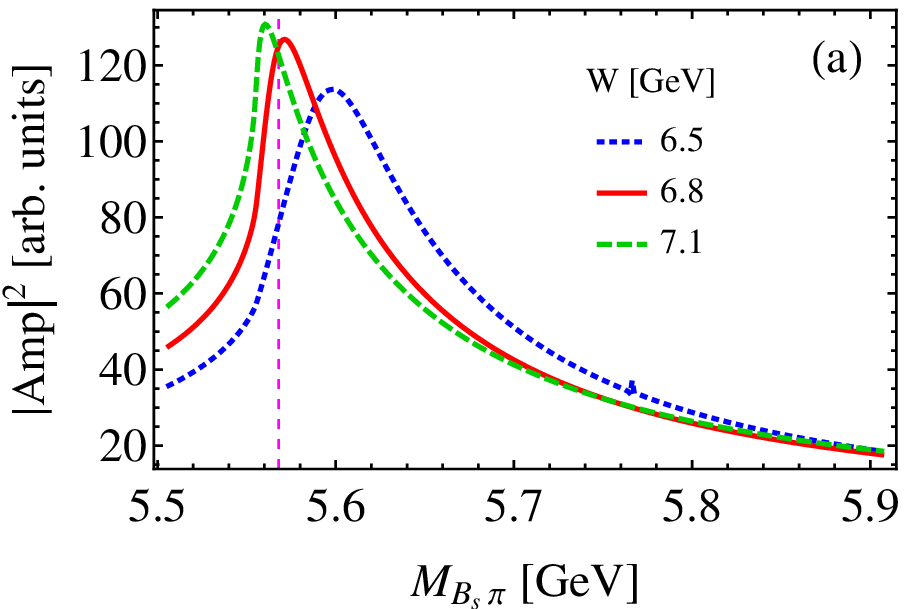}
	\includegraphics[width=0.38\hsize]{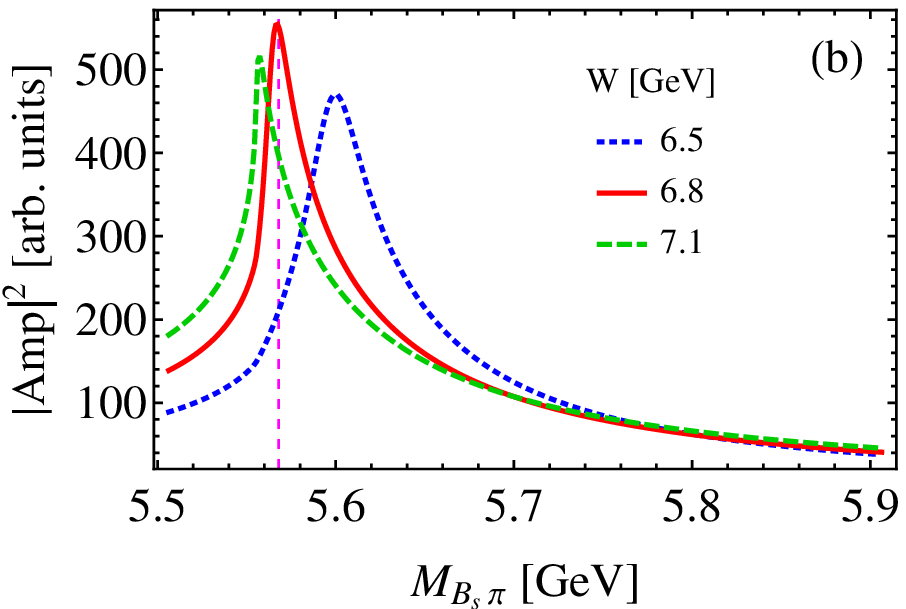}
	\caption{$M_{B_s\pi}$-dependence of the rescattering amplitude squared $|\mbox{Amp}|^2$. Plots (a) and (b) correspond to Figs.~\ref{feynmanBox}(a) and (b), respectively. $W$ is the invariant mass $\sqrt{s_1}$ of incident state $A$. The vertical dashed line indicates the position of $X(5568)$.}\label{distributionBox}
\end{figure}

Taking into account the above arguments, the triangle diagram in Fig.~\ref{feymanTri} can be changed into the box diagram in Fig.~\ref{feynmanBox} accordingly, and the numerical results for the $B_s\pi$ invariant mass distributions are displayed in Fig.~\ref{distributionBox}. Because the masses of $B_s$ and $B_s^*$ are different, it can be judged that there will no infrared divergence in these box diagrams~\cite{Ellis:2007qk}.  
Compared with the resonance-like bumps in Fig.~\ref{distributionTri}, the bumps in Fig.~\ref{distributionBox} are much narrower and more like resonance peaks. This implies that the interaction details of the scattering $B_s^*\pi\to B_s\pi$ may affect the TS mechanism to some extent.  

If the long range EM interactions play a dominant role in $B_s^*\pi\to B_s\pi$, one cannot expect there will be TS peaks arising in the $B_s\pi^0$ invariant mass distributions, because there is no $\gamma \pi^0\pi^0$ vertex. We can further conclude that if the observation of $X(5568)$ is due to the rescattering effects and the long range EM interaction dominates the scattering $B_s^*\pi\to B_s\pi$, there will no charge neutral partner of $X(5568)$. 

\subsection{Weak interaction process}
As stated before, the process $A\to B_s^* V$ can also happen via the weak interactions, such as $B_c^{(**)}\to B_s^* V$. Interestingly, according to the quark model calculation in Ref.~\cite{Godfrey:2004ya}, it can be noticed that there are many charm-beauty mesons $B_c^{(**)}$, of which the masses just fall into the region $ 6.2-7.5 $ GeV. This energy region has a large overlap with the TS kinematic region displayed in Table~\ref{KMregion}. This is another support that the TS mechanism may play a role in the observation of $X(5568)$.

\section{Summary}

In this paper, we investigated the invariant mass distributions of $B_s\pi$ via different rescattering processes. Because the TS's of rescattering amplitudes could be present for a very broad incident energy region, one can expect that the TS peaks may mimic the resonance-like structure $X(5568)$. The TS mechanism is highly process-dependent, which is different from other dynamic mechanisms. If the kinematic conditions of the TS are not fulfilled, there will be no peaks arising in the amplitudes. However, one would expect that the genuine particles should also appear in the processes where kinematic conditions of the TS are not fulfilled. If the observation of $X(5568)$ is due to the rescattering effects, because of the $P$-wave scattering characteristic of the process  $B_s^*\pi\to B_s\pi$, the quantum numbers of ``$X(5568)$'' would be $J^P=1^-$.
This is different from the assignment of diquark-antidiquark picture \cite{Agaev:2016mjb,Wang:2016mee,Wang:2016tsi,Chen:2016mqt} or $B^{(*)}\bar{K}$ molecular state model \cite{Xiao:2016mho}, where the quantum numbers are set to be $0^+$ or $1^+$. Further experiments with angular distribution analysis may help us to clarify the ambiguities and check different mechanisms.

\subsection*{Acknowledgments}
Helpful discussions with Makoto Oka, Wei Wang, Qiang Zhao and Shi-Lin Zhu are gratefully acknowledged.
This work is supported in part by the Japan Society for the Promotion of Science under Contract No. P14324, the JSPS KAKENHI (Grant No. 25247036), the National Natural Science Foundation of China (Grant No. 11275113)  and the Natural Science Foundation of
Shandong Province (Grant No. ZR2015JL001).

\end{document}